\documentclass[%
reprint,
superscriptaddress,
amsmath,amssymb,
aps,
pre,
floatfix,
]{revtex4-2}
\usepackage{graphicx}
\usepackage{dcolumn}
\usepackage{bm}
\usepackage{hyperref}

\usepackage{graphicx,setspace} 

\usepackage{bbm}
\usepackage{physics}
\usepackage{nicematrix}
\hypersetup{colorlinks=true, allcolors=blue}
\usepackage{amsmath} 
\usepackage{amssymb}
\usepackage{lineno}
\usepackage{hyperref}

\begin{document}

\title{Segmentation of dense and multi-species bacterial colonies using models trained on synthetic microscopy images}

\author{Vincent Hickl}
\email{vincent.hickl@empa.ch}
\affiliation{Laboratory for Biointerfaces, Empa, Swiss Federal Laboratories for Materials Science and Technology, Lerchenfeldstrasse 5, 9014 St. Gallen, Switzerland}
\affiliation{Center for X-ray Analytics, Empa, Swiss Federal Laboratories for Materials Science and Technology, Lerchenfeldstrasse 5, 9014 St. Gallen, Switzerland}
\affiliation{Laboratory for Biomimetic Membranes and Textiles, Empa, Swiss Federal Laboratories for Materials Science and Technology, Lerchenfeldstrasse 5, 9014 St. Gallen, Switzerland}

\author{Abid Khan}
\affiliation{Department of Physics, University of Illinois Urbana-Champaign, Urbana, IL, United States 61801}
\affiliation{NASA Ames Research Center, Moffett Field, CA, 94035, USA}

\author{René M. Rossi}
\affiliation{Laboratory for Biomimetic Membranes and Textiles, Empa, Swiss Federal Laboratories for Materials Science and Technology, Lerchenfeldstrasse 5, 9014 St. Gallen, Switzerland}

\author{Bruno F. B. Silva}
\affiliation{Laboratory for Biointerfaces, Empa, Swiss Federal Laboratories for Materials Science and Technology, Lerchenfeldstrasse 5, 9014 St. Gallen, Switzerland}
\affiliation{Center for X-ray Analytics, Empa, Swiss Federal Laboratories for Materials Science and Technology, Lerchenfeldstrasse 5, 9014 St. Gallen, Switzerland}
\affiliation{Laboratory for Biomimetic Membranes and Textiles, Empa, Swiss Federal Laboratories for Materials Science and Technology, Lerchenfeldstrasse 5, 9014 St. Gallen, Switzerland}

\author{Katharina Maniura-Weber}
\affiliation{Laboratory for Biointerfaces, Empa, Swiss Federal Laboratories for Materials Science and Technology, Lerchenfeldstrasse 5, 9014 St. Gallen, Switzerland}


\keywords{Single-cell segmentation, Bacterial self-organization, Machine learning, Image analysis, Active matter}


\begin{abstract}

The spread of microbial infections is governed by the self-organization of bacteria on surfaces. Limitations of live imaging techniques make collective behaviors in clinically relevant systems challenging to quantify. Here, novel experimental and image analysis techniques for high-fidelity single-cell segmentation of bacterial colonies are developed. Machine learning-based segmentation models are trained solely using synthetic microscopy images that are processed to look realistic using state-of-the-art image-to-image translation methods, requiring no biophysical modeling. Accurate single-cell segmentation is achieved for densely packed single-species colonies and multi-species colonies of common pathogenic bacteria, even under suboptimal imaging conditions and for both brightfield and confocal laser scanning microscopy. The resulting data provide quantitative insights into the self-organization of bacteria on soft surfaces. Thanks to their high adaptability and relatively simple implementation, these methods promise to greatly facilitate quantitative descriptions of bacterial infections in varied environments.

\end{abstract}

\maketitle

\section*{Introduction}

Microbial attachment and aggregation at surfaces are fundamental to the resilience of bacterial infections. After planktonic bacteria settle onto a substrate, they form microcolonies of small numbers of cells that become precursors for the biofilms \cite{QuispeHaro2024,Duvernoy2018} that make infections dramatically more resistant to medical intervention. The self-organization of these bacteria during the early stages of infection plays a crucial role in determining the progression of the infection \cite{Shimaya2022,You2018,Langeslay2023,Yan2016,Singh}. Understanding how bacterial attachment and microcolony formation vary across different surfaces reveals strategies to prevent infections before they fully develop~\cite{Pellegrino2022,Straub2022, Pan2022}. In turn, understanding the mechanics of biofilm architecture at the single-cell scale is essential to understanding how to disrupt their proliferation and create novel therapies that penetrate the bacteria's natural defenses. 

Novel imaging and analysis methods are needed to study bacterial self-organization in clinically relevant systems. The importance of single-cell segmentation for understanding the development of bacterial colonies is well understood, and some recent studies have achieved accurate segmentation of dense bacterial colonies and biofilms \cite{Cutler2022,Ollion2024,Jelli2023,Spahn2022}. However, to obtain optimal imaging conditions, the colonies in these studies are grown directly on glass coverslips, which does not correspond to the far more complex environments bacteria encounter in clinical settings. Developing methods for image analysis at surfaces with different geometric, material, and chemical properties has been a crucial challenge \cite{Wong2021,Muller2016}.

In principle, state-of-the-art segmentation methods using machine learning can overcome this problem, but typically require tedious human annotation of large image sets for training (meaning many cell labels must be drawn by hand in each image). As this problem transcends the study of bacteria and is relevant to a number of different disciplines, some current research in the field of image segmentation is focused on producing training datasets without human annotation. One promising approach is the use of artificial images that resemble experimental images closely enough to train segmentation models  \cite{Khan2023,Mill2021,Pawowski2022}. While some studies have applied this approach to bacterial colonies and biofilms, they require complex, explicit computational models of cell growth and of the imaging apparatus, experimentally measured point-spread functions, or extensive post-processing, making them challenging to apply to a broad range of experimental conditions \cite{Zhang2022, Toma2022, Hardo2022}. Simple, adaptable, and accessible computational tools are sorely needed to empower researchers to develop segmentation models tailored to their specific experimental system.

Existing studies of bacterial self-organization and novel methods for single-cell segmentation have mostly dealt with colonies consisting of a single species \cite{You2021,Singh,Zhang2022,Ollion2024}. However, the vast majority of real infections involve multiple species that collaborate or compete in complex ways \cite{Davies2003,Fazli2009}, and have profound effects on the efficacy of different treatments \cite{Camus2021,Beaudoin2017,Dittmer2023,Mill2021}. While species can be distinguished by using mutant strains expressing different fluorescent proteins, or through the use of different fluorescent stains, these methods greatly complicate experimental protocols and limit their relevance to clinical settings. The mechanical interactions between multiple bacterial species at the microscale remain poorly understood, partially because there are few segmentation models that can distinguish between strains based on morphology alone~\cite{Rombouts2022,Panigrahi2021,Zhang2020}. It is thus crucial that novel image segmentation methods be developed to include the capacity for multi-species segmentation.

Here, we present a new method for creating single-cell segmentation models from synthetic microscopy images produced through image-to-image translation. Using a custom microfluidic device, dense monolayers of rod-shaped bacteria are grown on PDMS films. In a different set of experiments, mixed suspensions of rod-shaped and spherical bacteria are imaged. Synthetic microscopy images of densely packed and multi-species bacterial colonies are produced using a simple and adaptable model. These synthetic images are then processed using a cycle generative adversarial network (cycleGAN)~\cite{Zhu2017} to resemble real images and serve as a training set for custom segmentation models. Thus, segmentation models adapted to a variety of experimental conditions can be trained quickly without human annotation. Dense monolayers of rod-shaped cells grown on soft substrates not optimized for high signal-to-noise ratios can be segmented with greater accuracy than with existing models from the literature. Quantitative information on the distribution of the bacteria can then be used to gain novel insights into bacterial self-organization. In images of mixed colonies, cells of different species can be automatically identified in datasets from both confocal and brightfield microscopy. This approach to analyzing bacterial colonies promises to greatly simplify the creation of accurate segmentation models tailored to a variety of in-vitro and in-vivo systems. 

\section*{Results}
\subsection*{A novel approach to bacterial segmentation with synthetic microscopy images and cycleGAN}

\begin{figure*}[t]
  \includegraphics[width=\linewidth]{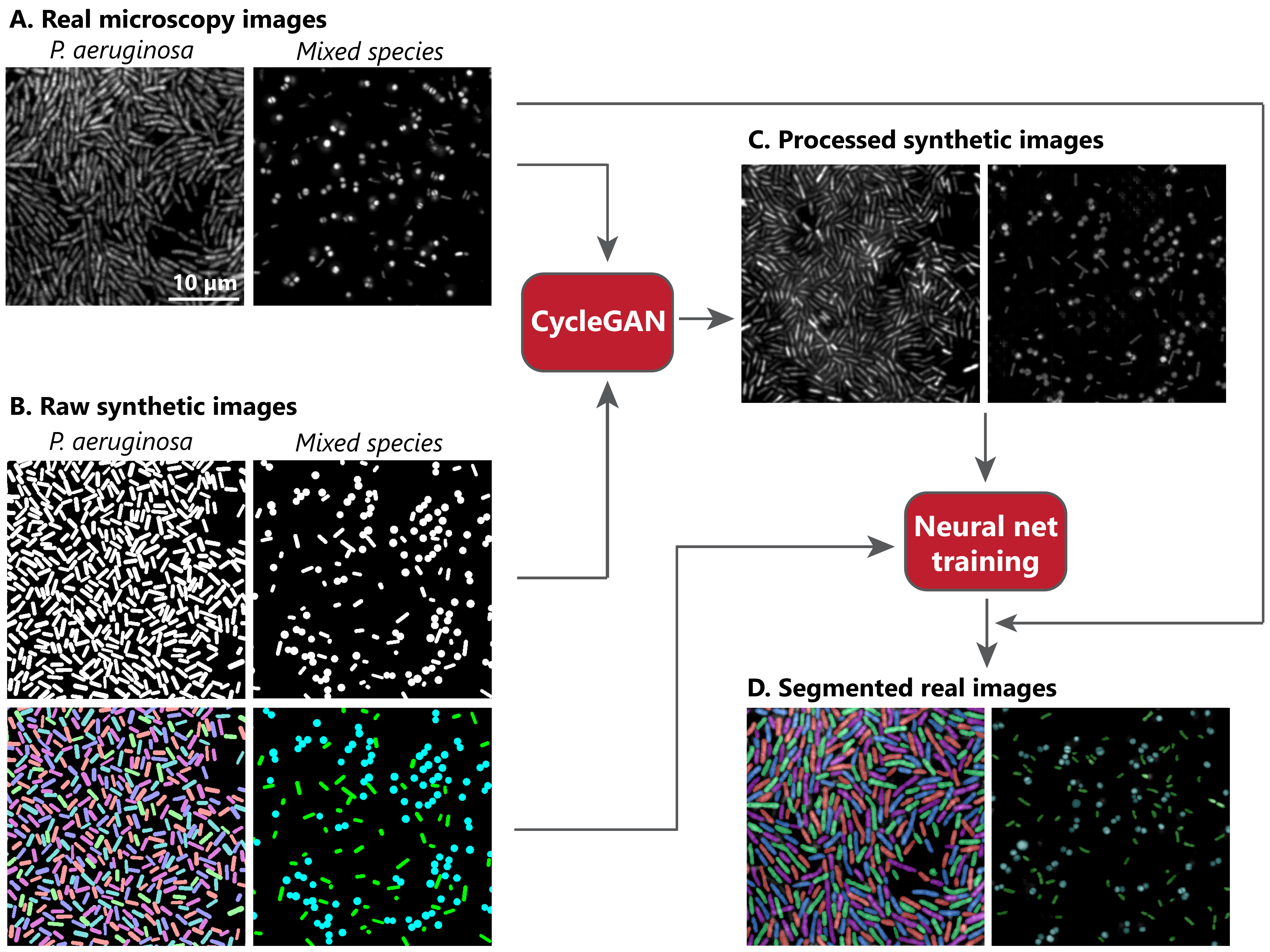}
  \caption{Workflow of segmentation model creation and application. (A) Real images of bacterial monolayers are taken using confocal microscopy. (B) Custom python algorithms are used to produce `raw' (binary) synthetic images along with ground truth masks. (C) Using cycleGAN, raw synthetic images are processed to qualitatively resemble real images. (D) The processed synthetic images and their masks are used to train a segmentation model in Omnipose~\cite{Cutler2022}. This model is then used to segment real images. The scale bar is $10$ \textmu m, and the scale is the same for all images.}
  \label{fig2:schematic}
\end{figure*}

The purpose of the methodology presented here is to provide an efficient and adaptable way to create image segmentation models in the life sciences, with a particular focus on microscopy applications for the study of bacterial infections. Our approach is summarized in Fig.~\ref{fig2:schematic}. First, real microscopy images of bacteria are recorded with custom imaging setups and multiple microscopy techniques. Then, custom computational models are used to create `raw' synthetic images of bacteria - images in which cell densities and shapes are approximately equal to those in the real images but that do not contain noise, blurred edges, anisotropic intensities, and other optical imperfections characteristic of real imaging techniques. Each synthetic image has an associated `mask' that encodes the location, morphology, and species of each cell. The real and raw synthetic images are then used as inputs for a cycle generative adversarial network (cycleGAN), which is an image-to-image translation method used here to `process' synthetic images by giving them optical characteristics to resemble the real images. Together with the original masks of the synthetic images, these processed synthetic images are then used to train neural networks to perform single-cell segmentation and species classification on real images.

\subsection*{Imaging dense single-species and mixed colonies of bacteria at interfaces}

\begin{figure*}
  \centering
  \includegraphics[width=0.9\linewidth]{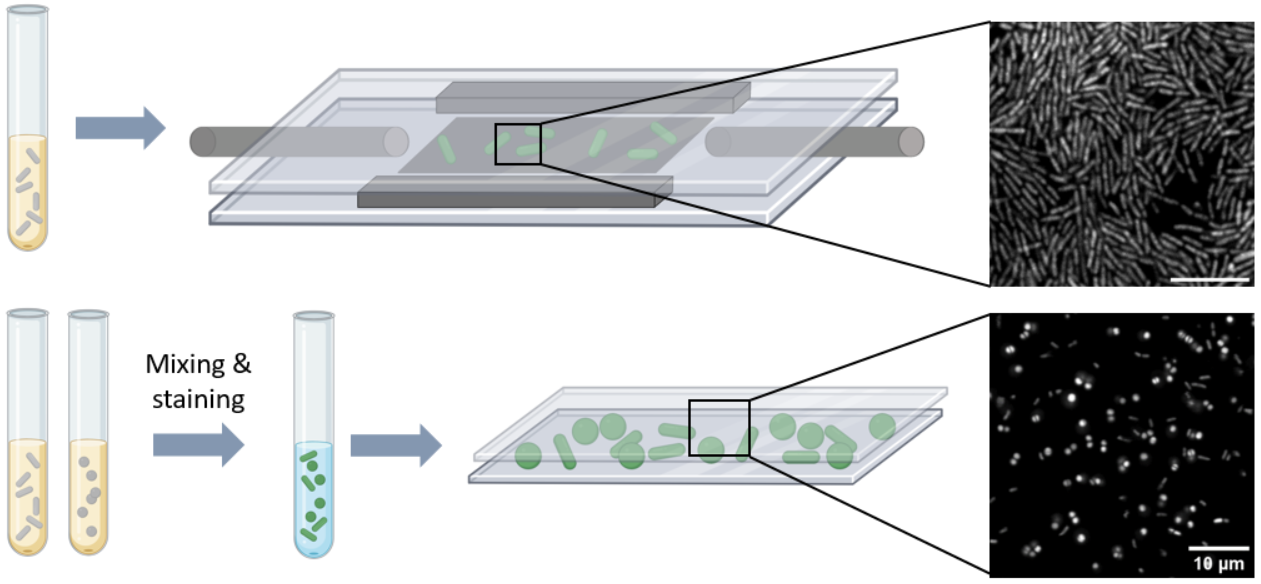}
  \caption{Experimental methods. Top: \textit{Pseudomonas aeruginosa} (\textit{P. a.}) are grown in liquid culture overnight, diluted in fresh medium, and seeded onto a thin PDMS film inside a custom-built microfluidic chamber. There, bacteria are grown overnight under a constant flow of medium with a low concentration of SYTO 9 to form a dense monolayer on the PDMS, which is then imaged using confocal microscopy. Scale bar is $10$ \textmu m. Bottom: \textit{Staphylococcus aureus} (\textit{S. a.}) and \textit{P. a.} are grown separately overnight, washed, resuspended in fresh PBS, mixed, and stained with SYTO 9. A small volume of the suspension is then placed between two coverslips and imaged using confocal microscopy. Scale bar is $10$ \textmu m. Figure created in part using BioRender.}
  \label{fig1:exp_meth}
\end{figure*}

To demonstrate the power and adaptability of our single-cell segmentation approach, experiments were conducted with colonies of clinically relevant bacterial strains using multiple sample preparation and imaging methods. Densely packed, single-species colonies of \textit{Pseudomonas aeruginosa} (\textit{P. a.}) were grown in a custom microfluidic device on $0.1$ mm thin films of PDMS (Fig.~\ref{fig1:exp_meth}, top). PDMS was chosen because it is an ideal model substrate used to study attachment and mechanosensing of bacteria at surfaces with different mechanical and geometric properties \cite{Straub2019,Pan2022,Pellegrino2022}. Bacteria were left to grow overnight in the chamber under a constant flow of medium supplemented with nucleic acid stain SYTO 9 for fluorescent imaging. The flow served to provide ample nutrients for continued growth and to wash away any cells not attached to the PDMS. The resulting dense monolayers were imaged through the PDMS (Fig.~\ref{fig1:exp_meth}, top right). PDMS has a different refractive index than the glass and the immersion oil used, and attenuates the incident and emitted light to and from the sample. Impurities within the PDMS may also interfere with sample illumination. These factors all contribute to a reduction in the signal-to-noise ratio and an increase in the point spread function (PSF) of the imaging system. These suboptimal imaging conditions (compared to bacterial colonies imaged directly on a glass cover slip) were chosen deliberately to develop a segmentation method that can provide quantitative information for a variety of experimental protocols involving bacteria at different surfaces.

To develop a method for simultaneous segmentation and classification of multiple strains, multi-species bacterial colonies with undifferentiated staining were formed. Liquid cultures of \textit{P. a.} and \textit{Staphylococcus aureus} (\textit{S. a.}) grown overnight were washed and resuspended in fresh PBS, mixed together, and either imaged immediately with brightfield microscopy or stained with SYTO 9 for confocal imaging (Fig.~\ref{fig1:exp_meth}, bottom). A droplet of this mixed suspension was then placed between two coverslips and imaged directly with a confocal microscope. Separately, unstained samples were similarly mounted for brightfield imaging. The resulting monolayers were less dense than the single-species monolayers of \textit{P. a.}, although clusters of cells could still be found (Fig.~\ref{fig1:exp_meth}, bottom right). Both rod-shaped cells (\textit{P. a.}) and spheroidal cells (\textit{S. a.}) can clearly be observed. These images are then used by an image-to-image translation algorithm to process synthetic images which in turn are used to train segmentation models.

\subsection*{Training segmentation model using image-to-image translation}

Synthetic microscopy images of both single-species and mixed bacterial colonies are created to provide training data for segmentation models without human annotation. Using custom python algorithms, rod-shaped bacteria (\textit{P. a.}) or mixtures of rod-shaped and spherical bacteria (\textit{S. a.}), are drawn as binary shapes on a dark background, forming `raw' synthetic images (Fig.~\ref{fig2:schematic}B). The cell density, degree of alignment (for rod-shaped bacteria), and degree of clustering were varied to provide a diverse dataset that includes the various cell configurations observed in the experimental images. \\

Using image-to-image translation in the form of a cycle generative adversarial network (cycleGAN) \cite{Lim2020}, synthetic images are processed to qualitatively resemble the real images without altering the ground truth of cell positions and orientations. Real experimental and synthetic images, created as described above, are used together to train a cycleGAN, which produces a generator that transforms synthetic images to closely resemble real experimental ones (Fig.~\ref{fig2:schematic}C).
Separate cycleGANs are trained for single-cell and mixed colonies. The synthetic images are modified to add noise, vary the brightness of the cells, and blur near cell boundaries or edges. However, the cells' positions and orientations are preserved, meaning the masks created for the synthetic images can later be used to train a segmentation model. 

From the processed synthetic images and corresponding masks, tailored segmentation models are created to segment and classify bacteria in each set of real images. Models are trained using the general image segmentation tool Omnipose \cite{Cutler2022}, which provides a framework to train deep neural network algorithms for bacterial segmentation. For our application, the training data consists of the synthetic images processed by cycleGAN along with their corresponding masks (Fig.~\ref{fig2:schematic}D). No real experimental images or other hand-annotated images were used for this training. Three different segmentation models were trained here: one for monolayers of \textit{P. a.} grown on PDMS, a second to identify and segment \textit{P. a.} in mixed colonies with \textit{S. aureus}, and a third for \textit{S. aureus} in the same mixed colonies. The latter two models, when used together, allow for simultaneous classification and segmentation of both species in mixed colonies without differential staining.

\subsection*{Segmenting dense \textit{P. aeruginosa} monolayers}

\begin{figure*}
  \includegraphics[width=\linewidth]{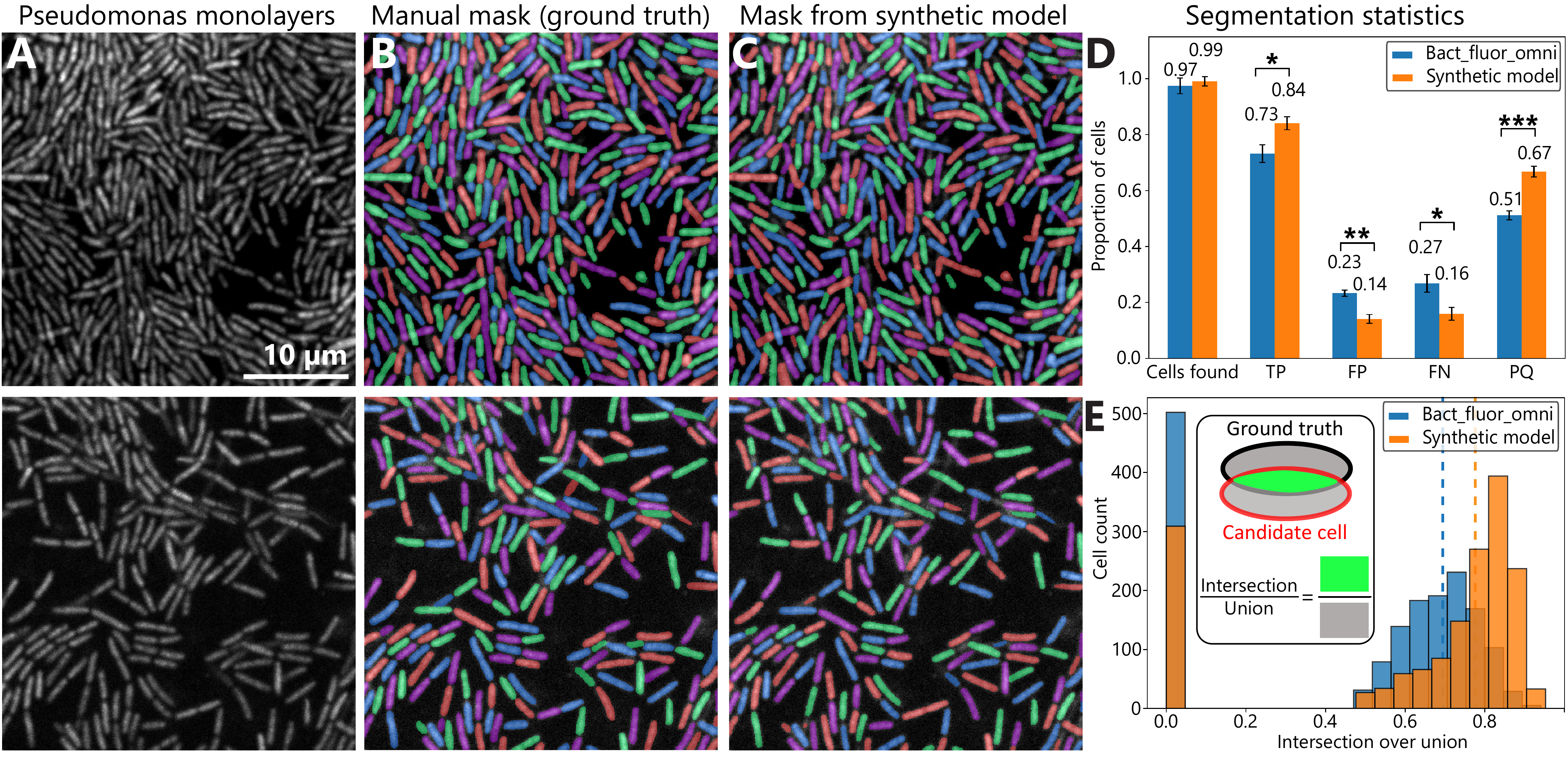}
  \caption{Segmentation of \textit{P. a.} monolayers. (A) Confocal microscopy images at $63\times$ magnification of dense (top) and dilute (bottom) monolayers of \textit{P. a.} stained with SYTO 9. The scale bar is $10$ \textmu m, and the scale is the same for all images. (B) Sample hand-annotated masks showing the ground truth positions of all bacteria in the monolayers. Colors are only used to visually distinguish cells and do not correspond to any physical parameters. (C) Sample masks produced by the segmentation model trained on synthetic microscopy images processed with cycleGAN (`synthetic model'). (D) Segmentation statistics and comparison to a segmentation model (`Bact\_fluor\_omni') trained on high-quality hand-annotated experimental images of bacteria \cite{Cutler2022}. True positives (TP), false positives (FP), and false negatives (FN) are given as a proportion of the number of cells in the ground truth mask, and panoptic quality (PQ)~\cite{Kirillov2018} is defined between $0$ and $1$. Error bars represent standard error from variation between images. The data are drawn from 13 different images such as the ones in panels (B) and (C), and include $1533$ cells in total. Asterisks represent significance from a t-test at the $P<0.05$, $0.01$, and $0.001$ levels, respectively. (E) Distribution of intersection over union (a common measure of how accurately cells are identified during segmentation) from both our model trained on synthetic images and Bact\_fluor\_omni. The bars at $IoU=0$ represent the number of false negatives. Dashed lines represent average IoUs. Inset: diagram of intersection over union for a model cell.}
  \label{fig3:pa_seg}   
\end{figure*}

Segmentation models trained on synthetic images of single-species colonies of \textit{P.a.} produce excellent single-cell segmentation of densely-packed colonies. Several experimental images, including densely packed and more dilute monolayers of cells (Fig.~\ref{fig3:pa_seg}A), were manually labeled to test the accuracy of this segmentation model. These manual labels provided `ground truth' masks to which masks from the segmentation model could be compared (Fig.~\ref{fig3:pa_seg}B). In total, $13$ images containing a total of $1533$ cells were manually annotated. In all test images, the masks generated by the model trained on synthetic images (`synthetic model') accurately reproduce the labels in the ground truth mask. Cells are accurately and reliably distinguished from the background, and nearby cells are distinguished from each other with few exceptions, while individual cells are rarely incorrectly divided into smaller ones (Fig.~\ref{fig3:pa_seg}C).

Several quantitative metrics are calculated to assess the quality of the segmentation model. These include the number of total cells found by the segmentation model (`candidate cells') as well as the number of true positives (TP), false positives (FP), and false negatives (FP), all as a proportion of cells in the ground truth mask (Fig.~\ref{fig3:pa_seg}D). The `panoptic quality' (PQ) is also computed, which is a common measure of overall segmentation quality that takes into account the proportion of cells correctly identified and how well the candidate cells match the corresponding cell in the ground truth \cite{Kirillov2018}. The number of cells identified by the synthetic model is, on average, $99\pm3$\% of the number of cells in the ground truth mask, suggesting neither under- nor oversegmentation. On average, $84$\% of cells are correctly identified (true positives). Additionally, the intersection over union (IoU) is computed for each candidate cell, which measures how well the candidate cell matches the ground truth (Fig.~\ref{fig3:pa_seg}E, inset). True positives are defined as candidate cells with $IoU > 0.5$. 

The segmentation model trained on synthetic images outperforms existing state-of-the-art segmentation models available in the recent literature. Here, our `synthetic model' is compared to `Bact\_fluor\_omni', a model trained in Omnipose using hand-annotated experimental images of various bacteria strains with different densities and morphologies. This model is optimized for high-quality fluorescent images and models like it have been shown to outperform other recent segmentation techniques~\cite{Cutler2022}. Across all calculated metrics, our model trained on synthetic images performs significantly better than Bact\_fluor\_omni. On average, the percentage of cells correctly identified is $11$ percentage points higher, the panoptic quality is $16$ percentage points higher, and the mean IoU of candidate cells from the synthetic model is $30$\% higher compared to those from the model trained on a high quality set of hand-annotated experimental images. 

\subsection*{Single-cell statistics from dense \textit{Pseudomonas} monolayers}

\begin{figure*}
  \includegraphics[width=\linewidth]{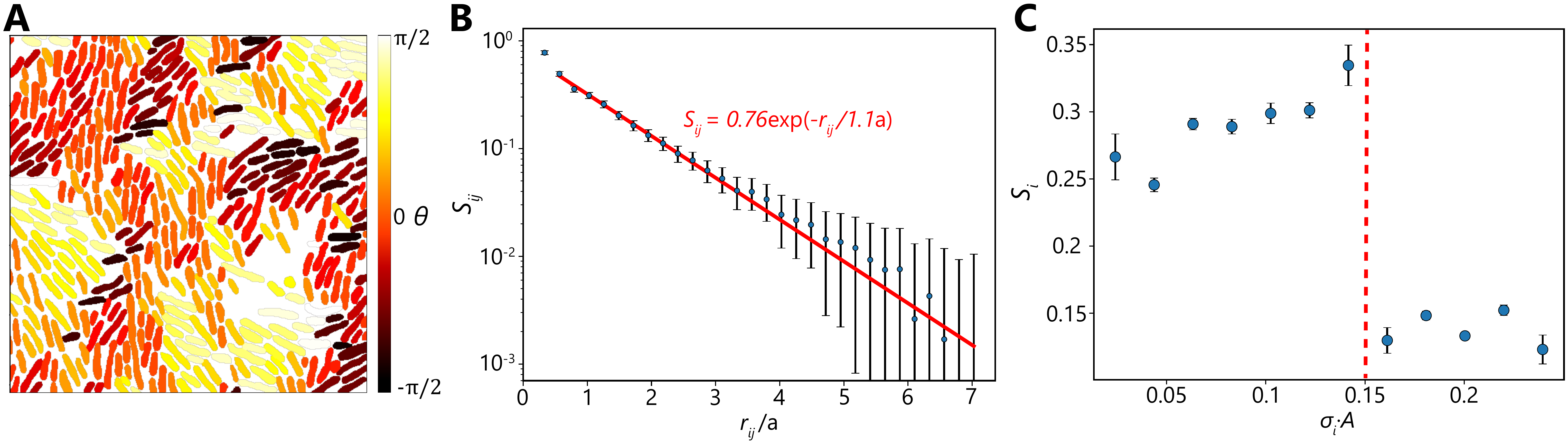}
  \caption{Single-cell statistics in dense \textit{Pseudomonas} monolayers (A) Representative region of densely packed \textit{P.a.}, segmented using a model trained on synthetic images and colored according to their angles relative to the vertical axis, showing parallel alignment of nearby cells. (B) Alignment of pairs of cells ($N=22149$), measured by the order parameter $S_{ij}$, as a function of their dimensionless separation, $r_{ij}/a$, where $a$ is the mean cell length. Exponential fitting curve is shown in red. Error bars represent variation between different cell pairs at each distance. (C) Local cell alignment ($N = 33300$) as a function of dimensionless packing fraction $\varphi_i\equiv\sigma_iA$, where $\sigma_i$ is the number of cells per area and $A$ is the mean cell area. The density for each cell is an average measured across an area of radius $10$ \textmu m around each cell. The local alignment for the $i^{th}$ cell, $S_i$, is the average of the order parameter $S_{ij}$ for all cell pairs ${i,j}$ that include the $i^{th}$ cell. The data analyzed here include images from $2$ separate experiments performed several weeks apart using the same protocol.} 
  \label{fig4:pa_stats}   
\end{figure*}

Accurate single-cell segmentation enables quantitative statistics about bacterial self-organization at the cellular scale to be determined. Using the masks provided by the segmentation model trained on synthetic images, the positions, orientations, and dimensions of all cells in densely packed monolayers of \textit{P. a.} are determined. Since the bacteria in this experiment have no front or back, they have $180^\circ$ symmetry, and their orientations are given by an angle $\theta$ in the interval $(-\pi/2,\pi/2]$ (Fig.~\ref{fig4:pa_stats}A). The parallel alignment of nearby cells can clearly be observed. Thus, single-cell segmentation provided precise information on how cells in dense colonies are distributed.  

Local cell alignment decays exponentially with a correlation length comparable to the length of a single cell. For any pair of cells separated by a distance $r_{ij} = \sqrt{(x_i - x_j)^2 + (y_i - y_j)^2}$, where $(x_i, y_i)$ is the position of the $i^{\text{th}}$ cell, the alignment is quantified via the order parameter
\begin{equation}
    S_{ij} = 2\cos^2(\theta_i - \theta_j) - 1\text{,}
\end{equation}
where $\theta_i$ is the angle of the $i^{th}$ cell measured relative to the vertical axis. Pairs of cells parallel/perpendicular to each other have $S_{ij} = \pm1$, respectively. The dimensionless cell separation is determined by dividing $r_{ij}$ by the mean cell length $a=2.8$ \textmu m. Calculating $r_{ij}/a$ and $S_{ij}$ for every pair of cells with $r_{ij}<20$ \textmu m in our images ($N=22149$) shows how the orientational order of the cells depends on distance (Fig.~\ref{fig4:pa_stats}B). We find that $S_{ij}$ decays exponentially as a function of $r_{ij}/a$. Fitting gives a correlation length of $0.9a$ or $2.5$ \textmu m (slightly shorter than the length of the average \textit{P. a} cell). 

Local cell alignment depends non-monotonically on local cell density. For each cell analyzed ($N = 33300$), the average local alignment to its neighbors (defined as all cells within twice the correlation length, $1.8a$ or $5$ \textmu m), $S_{i}$, and the dimensionless local cell density, or packing fraction, $\varphi \equiv \sigma_iA$ are calculated. Here, $\sigma_i$ is the number of cells per unit area within a $10$ \textmu m radius of cell $i$, and $A$ is the average cross-sectional area of all cells, $1.65 \text{ \textmu m}^2$. The packing fraction varied from $0.01$ to $0.25$, and the local cell alignment varied from $-0.77$ to $0.90$. For low packing fractions ($\varphi < 0.15$), cell alignment increases with increasing cell density. However, a sharp drop in cell alignment occurs at $\varphi \approx 0.15$, above which no significant dependence of cell alignment on packing fraction could be observed (Fig.~\ref{fig4:pa_stats}C). Results such as these require individual cell dimensions, positions, and orientations to be measured simultaneously in dense colonies, which is only possible here thanks to the accurate segmentation provided by our model trained on synthetic images. 

\subsection*{Segmentation and classification of multi-species colonies}

\begin{figure*}[h]
  \includegraphics[width=\linewidth]{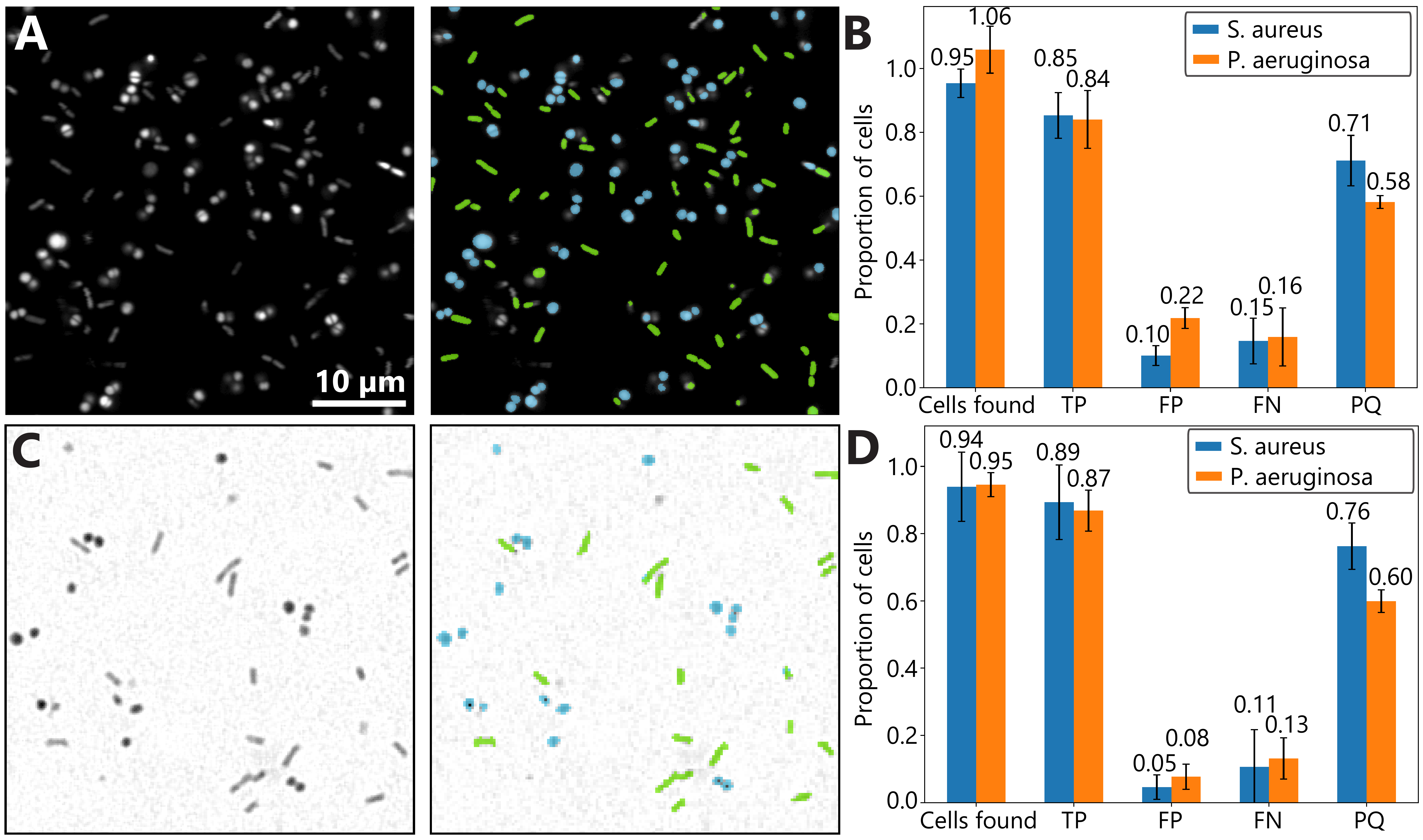}
  \caption{Simultaneous segmentation and classification of \textit{P. aeruginosa} and \textit{S. aureus}. (A) Confocal microscopy image ($63\times$) of a mixed bacterial suspension stained with SYTO 9. In the right panel, green and blue cells represent individual bacteria classified as \textit{P. a.} and \textit{S. a.}, respectively, by a model trained on synthetic microscopy images. The scale bar is $10$ \textmu m, and the scale is the same for all images. (B) Segmentation statistics for multi-species segmentation of confocal microscopy images. True positives (TP), false positives (FP), and false negatives (FN) are given as a proportion of the number of cells in the ground truth mask, and panoptic quality (PQ)~\cite{Kirillov2018} is defined between $0$ and $1$. Error bars represent standard error from variation between images. The data are drawn from $4$ different images, such as the ones in panel (A), and include $510$ cells in total. (C) Brightfield microscopy image ($40\times$) of a mixed bacterial suspension. In the right panel, green and blue cells represent individual bacteria classified as \textit{P. a.} and \textit{S. a.}, respectively, by a model trained on synthetic microscopy images. (D) Segmentation statistics for multi-species segmentation of images from brightfield microscopy. Statistics are given as a proportion of the number of cells in the ground truth mask. Error bars represent standard error from variation between images. The data are drawn from 6 different images, such as the ones in panel (C), and include $233$ cells in total.} 
  \label{fig5:mixed_seg}   
\end{figure*}

Segmentation models trained on synthetic images can achieve single-cell segmentation and classification of multi-species colonies. Suspensions of \textit{P. a.} and \textit{S. a.} are mixed and stained with SYTO 9 as described above, and imaged using a confocal laser scanning microscope. Two separate segmentation models are trained in Omnipose on synthetic images processed by cycleGAN, and combined in a single Python script to identify both species (Fig.~\ref{fig5:mixed_seg}A). On average, the model accurately identifies and segments $85$\% and $84$\% of cells for \textit{S. a.} and \textit{P. a.}, respectively (Fig.~\ref{fig5:mixed_seg}B). Neither model systematically under- or oversegments cells, on average.

Multi-species segmentation is also possible for unstained mixed colonies imaged using brightfield microscopy. Mixed suspensions are simply deposited on a coverslip and imaged with no staining or special sample preparation (Fig.~\ref{fig5:mixed_seg}C). A separate set of segmentation models is trained for this case from the ones used for stained bacteria. For \textit{S. a.} and \textit{P. a.}, the model accurately identifies $89$\% and $87$\% of cells, on average, respectively, which is slightly better than the segmentation performed on confocal images (Fig.~\ref{fig5:mixed_seg}D). Neither of these models systematically over- or undersegments cells. These results demonstrate both the power and adaptability of segmentation models trained on synthetic images, as these can easily be trained for a variety of imaging techniques and sample preparation methods.  

\section*{Discussion}

Using synthetic microscopy images processed with cycleGANs, we show that single-cell image segmentation models can be efficiently created for a variety of experimental setups and imaging methods without tedious manual annotation. Accurate cell segmentation and classification is achieved even when cells are densely packed (that is, cells are touching or overlapping), when multiple species of different shapes are mixed, and when bacteria are grown on substrates that are not optimized for high-resolution imaging. This approach has several advantages over other ways to create segmentation models. First, it does not require any human annotation of training images, which is not only very time-consuming but can also introduce human biases into the training data \cite{Thiermann2024,Geiger2021}. Second, the use of image-to-image translation greatly simplifies the creation of synthetic images suitable for the training of segmentation models. The only knowledge required for this step is the approximate geometry and concentration of cells in the images; no excplicit modeling of the biology, physics, or optics of the experiment is required. The addition of noise, blurring (mimicking the point spread function), and small corrections to shape, size, and morphology are automatically handled by the cycleGAN. Third, and perhaps most importantly, these features of our approach make it exceedingly efficient to create highly specialized segmentation models for different experimental setups. In general, this adaptable method could be used for a wide variety of image analysis applications in the life sciences and beyond. 

The ability to accurately perform single-cell segmentation in different environments is important to understand the self-organization of bacteria and, in turn, shed light on the properties of dangerous biofilms. In particular, it enables the simultaneous determination of cell positions, orientations, and morphologies, as demonstrated here. Segmentation of dense \textit{P. a.} monolayers revealed quantitative information about their self-organization. The correlated alignment between nearby cells was found to decay exponentially over distances up to $~7$ times the average cell length. Surprisingly, local cell alignment was not found to increase monotonically with cell density. This result challenges conventional descriptions of orientational order in bacterial colonies, according to which cell alignment is driven by steric interactions that arise when cells are densely packed \cite{You2021,Langeslay2023}. Such alignment would also be expected from the Onsager theory of liquid crystals, which predicts an isotropic-to-nematic transition as the concentration of hard rods in a system increases~\cite{Onsager1949}. Instead, we find here that there is a sharp drop-off in local cell alignment above a packing fraction of $\varphi\approx0.15$. Further work using single-cell segmentation of bacterial colonies under different conditions is required to determine how surface properties, cell motility, or other factors affect their self-organization. An important extension of the work presented here is to develop similarly efficient segmentation models in 3D and investigate the effects of more complex geometries on bacterial colonies. Work on this problem is currently underway and will enable the analysis of bacterial aggregates in more varied and relevant settings. 

The methods presented here provide ways to address several key challenges related to single-cell segmentation. As previous studies have shown, the precise morphology and size of cell masks in the training data is extremely important to make sure cell dimensions can be accurately measured after segmentation \cite{Thiermann2024}. The use of synthetic images gives precise control over the shape of cells in the training data and removes human biases that may occur during manual annotation. The range of cell dimensions in synthetic images can be made wider than that observed for real cells, ensuring the resulting segmentation model is sensitive to the full range of cell sizes in the real images. The importance of cell shape is underscored by our results on multi-species segmentation in confocal images. While \textit{S. a.} cells were modeled as perfect circles, they often appeared more oval-shaped or amorphous in real images (particularly during cell division), leading to a higher rate of false positives from the model trained to identify non-circular cells. Future models will address this limitation by giving \textit{S. a.} more diverse shapes in the synthetic images. Thanks to the high-quality image-to-image translation provided by cycleGAN, such changes to the synthetic images are relatively easy to implement, meaning the slightly higher false positive rate associated with the \textit{P. aeruginosa} model for confocal microscopy images of mixed colonies is not a fundamental limitation of our approach. In fact, we expect this methodology to facilitate the segmentation of more species with complex shapes, which has been a key challenge in the study of the spatiotemporal development of biofilms \cite{Wong2021}.

In real infections, bacterial species rarely act alone, and the interactions between multiple species must be better understood in order to develop effective new therapies, particularly alternatives to antibiotics. Single-cell, multi-species segmentation is an important component of this challenge, as it is necessary for understanding mechanical interactions in realistic microbial communities. When designing experiments to investigate bacterial infections in clinically relevant systems, it is generally desirable to avoid the addition of multiple fluorescent stains during bacterial growth or the use of fluorescent mutant strains since these are not present in vivo. Staining both species simultaneously at the end of an experiment is comparatively less likely to introduce confounding effects but makes differential staining of different species impractical. Thus, the ability to distinguish multiple species that are either unstained or stained with a single fluorescent marker is highly useful. Species classification tools have already been developed to identify bacterial strains in images with great accuracy for high-throughput identification of pathogens~\cite{Kang2020,Kim2023}. However, these existing methods only work when the images contain a single species. We are convinced that the methods developed here will pave the way towards more versatile tools that can be applied to a wide range of clinically relevant scenarios.

For all of our models, the final quality of the segmentation can be further improved by introducing additional post-processing steps to the masks produced by the segmentation model. Cell masks can be filtered by size, shape, and position in a variety of ways depending on the experimental system or the desired information to be extracted. For example, principal component analysis can be used to categorize cell masks by their shape and isolate ones that were likely segmented incorrectly for further post-processing \cite{Zhang2022}. Additionally, if segmentation is combined with tracking, additional metrics become available to correct potential segmentation errors \cite{Ollion2024}. It should be noted that the segmentation framework used here, Omnipose, was not designed for the purpose of multi-species classification. It is possible that a different network architecture could provide better results with the training data produced here. These post-processing steps and the development of more specialized tools for segmenting mixed bacterial colonies lie beyond the scope of this work, whose primary purpose is to show that synthetic microscopy images processed by cycleGAN provide an excellent replacement for hand-annotated training data and greatly increase the efficiency with which segmentation models are created. This approach is highly synergistic with other recent advances in bacterial segmentation, which stand to benefit from the rapid creation of training data without complex modeling, as demonstrated here.

\section*{Materials \& Methods}

\subsubsection*{Bacteria cultures on PDMS in microfluidic device}

Single-species cultures of \textit{Pseudomonas aeruginosa} (\textit{P. a.}) were grown on top of thin PDMS sheets in a microfluidic device. \textit{P. a.} MPAO1 \textit{flgE} knockout mutants, constructed as previously described \cite{Valentin2022}, were grown overnight in 30\% tryptic soy broth (TSB) with 0.25\% glucose. The culture was diluted to $OD=0.2$, and then diluted $50\times$ in fresh medium.

To construct a flow chamber, thin PDMS sheets were formed by pouring 0.55 g of Sylgard 184 at a base to curing agent ratio of 10:1 into a $8.5$ cm diameter petri dish. After degassing in a vacuum chamber, the sheets were cured for 2 hrs at $80 ^{\circ}$C, resulting in sheets of thickness approximately 100 \textmu m. A $2\times1$ cm piece of the resulting film was then cut out and placed on a \#1.5 coverslip. On either side of this PDMS film, two pieces of $1$ mm thick PDMS (formed using the same procedure as the thin films) were placed on the coverslip as spacers, and polyethylene tubes with an outer diameter of $1$ mm (Huberlab), were attached to the coverslip on either side of the PDMS film. Then, $0.5$ mL of the diluted \textit{P.a.} was pipetted onto the thin PDMS, seeding the surface with some attached cells for subsequent growth. Finally, another coverslip was placed on top of the spacers, and the edges were sealed with epoxy, forming a sealed microfluidic flow chamber for controlled growth of bacteria on the PDMS film. 

Bacteria were grown on the PDMS overnight under constant flow ($0.01$ mL/min) from a syringe pump of fresh growth medium supplemented with $1$ \textmu M SYTO 9 nucleic acid stain. The steady addition of SYTO 9 at a low concentration ensured that all bacteria were stained while preserving their viability. The resulting dense monolayers of \textit{P.a.} were then imaged using a Zeiss LSM780 confocal microscope with a $63\times$ oil immersion objective. 

\subsubsection*{Multi-species staining and imaging}

To test the simultaneous segmentation of multiple species, \textit{P. a.} and \textit{Staphylococcus aureus} (\textit{S. a.}) strain ATCC 6538 were cultured overnight as described above. The cultures were separately centrifuged at $7000$g for 10 minutes, and the pellets were resuspended in phosphate-buffered saline (PBS). For confocal imaging, some samples were stained with $2.5$ \textmu M SYTO 9. Equal volumes of both suspensions were mixed together via vortexing, and $5$ \textmu L of the mixture was placed between two coverslips for imaging. Confocal imaging was performed as described above, and brightfield imaging was performed using a Nikon Eclipse Ti2 microscope at $40\times$ magnification.

\subsubsection*{Synthetic images for segmentation model training}

Synthetic images of bacteria at interfaces were created using custom programs written in Python, as described in the Supporting Information (SI). In brief, to model rod-shaped \textit{P. a.} cells, bright rectangles with circular caps were drawn with various positions and orientations on a dark background. In some images, cells were drawn with random orientations, while in others, they were aligned parallel to their nearest neighbors to simulate the tendency of bacteria to exhibit orientational order. The degree of alignment was varied by introducing a random noise term to the angle of each cell. Images with varying degrees of alignment were produced in each dataset to ensure the training data for the segmentation model were not biased towards highly aligned or randomly aligned cells. \textit{S. a.} cells were modeled as circular disks. In some images, cells were drawn in clusters to include regions of more densely packed and more dilute cells. Additionally, a maximum overlap parameter was implemented to ensure neighboring cells could touch but did not overlap excessively. Cell dimensions were chosen to match those in a randomly chosen, hand-measured sample of cells from real experimental images. Similarly, the cell density in synthetic images was chosen to roughly match the density observed in experimental images.

\subsubsection*{Processing synthetic images using cycleGAN}
We train a cycleGAN to use one of its components: the generator that inputs a synthetic image and outputs a processed synthetic image. Training a cycleGAN requires two datasets $X$ and $Y$ that we want to translate between. The network consists of four models: two discriminators and two generators. The generators $G_Y:X\to Y$ and $G_X:Y\to X$ are supposed to take an image $x, y$ from one domain and process it to resemble an image belonging to the other domain $G_Y(x), G_X(y)$. The discriminators $D_X:X\to \mathbb{R}$, $D_Y:Y\to\mathbb{R}$ are meant to take an image $\hat{x}, \hat{y}$ and output whether it belongs in their respective domains $D_X(\hat{x}), D_Y(\hat{y})$. The two generators and two discriminators train by competing with each other: the generators are trained to fool the discriminators, and the discriminators are trained to tell apart generator images from real images. 

The model architectures and losses used in this work are the same that are used in \cite{Khan2023}. The training set of cycleGAN is composed of experimental and synthetic bacterial images. To prepare the cycleGAN training set, for each data set, we normalize all the image intensities to [-1, 1] within their group. We then cut them into $256\times256$ patches. The batch size is 42, however when adding each patch into the batch, they are randomly augmented by a combination of rotation (multiples of 90 degrees) and flipping (vertically or horizontally). We add Gaussian noise (std = 0.1) to the initially noise-free synthetic images to ensure enough variability across the simulation data set. All networks were trained for 400 epochs, where in the first 200 epochs, the learning rate of the generators was $2\times 10^{-5}$, and in the last 200 epochs, the learning rate linearly decayed to zero. In the training procedure, it is important for the discriminators not to be too ahead of the generators so that the generators can train better against the discriminators. To ensure this, we set the learning rate of the discriminators $\eta_{\text{disc}}$ to be varied at each training step,
\[\eta_\text{disc} = 2\eta_\text{gen}\left|1-\alpha_{\text{disc}}\right|\]
where $\eta_\text{gen}$ is the learning rate of the generators, and $\alpha_{\text{disc}}$ is the accuracy, or fraction of the discriminators correctly evaluating test images. 

It is not always true that the trained model at the last epoch provides generated images that most closely resemble the real image dataset. To obtain the best model in our training process, we construct a dataset of processed synthetic images at every 10 epochs and compare it to the dataset of real images. This is done using the Fr\'echet inception distance (FID) between the real dataset and the processed synthetic dataset produced at a given epoch~\cite{Heusel2017}. We choose the model whose processed image dataset gives the lowest FID relative to the real dataset.

\subsubsection*{Segmentation model training}
Segmentation models were trained using the synthetic microscopy images generated by cycleGAN using the Omnipose package for Python \cite{Cutler2022}. This package is open-source and freely available (\url{github.com/kevinjohncutler/omnipose}, installed development version April 30th, 2024). Details and training parameters can be found in the SI. The model trained on real, hand-annotated images of bacteria against which the performance of our algorithm was compared (`Bact\_fluor\_omni') is included when installing Omnipose.

For the model trained to segment monolayers of \textit{Pseudomonas aeruginosa} (\textit{P. a.}), the training data consisted of $126$ processed synthetic images of dimensions $512\times512$ pixels and their corresponding masks. For the models trained to segment cells in mixed colonies of \textit{P. a.} and \textit{Staphylococcus aureus} (\textit{S. a.}) imaged using confocal microscopy, the training data consisted of $226$ processed synthetic images of dimensions $256\times256$ pixels. For the models trained to segment cells in mixed colonies imaged using brightfield microscopy, the training data consisted of $441$ processed synthetic images of dimensions $256\times256$ pixels. 

For segmentation of multi-species colonies, the synthetic images used were the same for the models trained to detect rods and circles, but in each case the corresponding masks only contained cells of a single species (rods or circles, respectively). The two models trained to segment \textit{P. a.} and \textit{S. a.} cells, respectively, were then combined in single Python program to perform simultaneous segmentation and classification. In most test images used here, a small number of cells was identified by both models. These cells were assigned to the species for which the false negative positive rate was lower overall in that set of test images. Here, for both confocal and brightfield microscopy, this means that cells identified by both models were classified as \textit{S. a.} by our program. 

\subsubsection*{Calculating segmentation model performance}
The performance of the segmentation models was evaluated by calculating several metrics that can then be used to compute a single parameter, ``Panoptic quality" (PQ), which ranges from $0$ to $1$ and provides a general metric for the quality of the segmentation \cite{Kirillov2018}. PQ is defined as 
\begin{equation}
    PQ = \frac{\sum_{(g,c)\in TP} IoU(g,c)}{|TP|}\times\frac{|TP|}{|TP|+\frac{1}{2}|FP|+\frac{1}{2}|FN|}\text{,}
\end{equation}
where $g$ and $c$ are ground truth and candidate cells, respectively, $IoU$ is the intersection over union of each ground truth and candidate pair, and the sum is taken only over pairs for which $IoU>0.5$, meaning the candidate is a true positive (TP). $|TP|$, $|FP|$, and $|FN|$ are the number of true positives (candidate cells from the segmentation model that correspond to actual cells), false positives (candidate cells that do not correspond to real cells), and false negatives (real cells not identified by the model). The first term in the $PQ$ is sometimes called ``segmentation quality" and measures how well true positives match with the corresponding real cell, on average. The second term is the ``recognition quality," which measures the model's overall ability to correctly find cells. Thus, $PQ$ provides a comprehensive metric for evaluating the performance of a segmentation model.

\subsubsection*{Quantifying cell alignment in \textit{P. a.} monolayers}
The results of the single-cell segmentation of densely-packed, rod-shaped \textit{P. a.} were used to extract quantitative information about the bacteria's self-organization. All analyses were performed with custom Python code using open-source and freely available libraries. The \verb|regionprops| function from the Sci-kit image analysis package was used to extract the positions, orientations (measured here so that the orientation angle $\theta=0$ when a rod-shaped cell is aligned with the vertical axis), and dimensions of each cell. Cells in contact with the edge of the image or below a size threshold of $10$ pixels were removed prior to further analysis. Additionally, when calculating local cell densities (see Fig.~\ref{fig4:pa_stats}C), only cells at least $10$ \textmu m from the nearest edge of the image were included in the analysis.

\medskip


\section*{Acknowledgements}
We would like to thank Dr. Qun Ren for fruitful discussions and scientific input, and her team for their support in the lab.

\section*{Data Availability}

All real and raw synthetic images used in the figures above and to train CycleGAN are freely available on Zenodo: 
\url{https://doi.org/10.5281/zenodo.12759488}.

The Python code to produce raw synthetic images and to reproduce the analysis in this article is available on GitHub: \url{github.com/vhickl/synth-bacteria-segmentation}.

All code required to train CycleGANs to process synthetic images, to calculate FID scores, and generate processed synthetic images using trained models is available on GitHub: \url{github.com/abid1214/cyclegan}.

\medskip

\bibliographystyle{apsrev4-2}

\bibliography{bibliography}

\end{document}